\pdfoutput=1
\documentclass[11pt]{article}

\usepackage[a4paper,margin=2.5cm]{geometry}
\usepackage{amsmath}
\usepackage{amssymb}
\usepackage{booktabs}
\usepackage{graphicx}
\usepackage{microtype}
\usepackage[numbers,sort&compress]{natbib}
\usepackage{siunitx}
\PassOptionsToPackage{hyphens}{url}
\usepackage[hidelinks]{hyperref}

\makeatletter
\g@addto@macro{\UrlBreaks}{\UrlOrds\do\/\do-\do_\do.\do:}
\makeatother

\graphicspath{{figures/main/}}
\setcitestyle{super,sort&compress}
\title{Local topology hallucinations in universal machine-learned interatomic potentials}
\author{%
Xiaoqing Liu$^{a}$, Yangshuai Wang$^{b,*}$, and Lei Shen$^{a,*}$\\[0.6em]
\small $^{a}$Department of Mechanical Engineering, National University of Singapore,
\\[-0.1em]\small 117575 Singapore\\
\small $^{b}$Department of Mathematics, National University of Singapore,
\\[-0.1em]\small 10 Lower Kent Ridge Road, Singapore\\[0.4em]
\small $^{*}$Corresponding authors:
\href{mailto:yswang@nus.edu.sg}{yswang@nus.edu.sg};\\
\href{mailto:shenlei@nus.edu.sg}{shenlei@nus.edu.sg}%
}
\date{}

\begin{document}
\begingroup\sloppy
\maketitle
\endgroup

\begin{abstract}
Predicting metastable states with universal machine-learned interatomic potentials requires that they preserve the minima and saddles of the reference potential-energy surface.
Here we identify local topology hallucination, in which distinct model minima relax into the same reference basin.
This failure arises from mode-selective curvature-sign errors shared by MACE, CHGNet and M3GNet.
Curvature signs predicted by two of the three models agreed with density-functional calculations in 55 of 57 positive but only 5 of 56 negative cases.
In LiGe$_2$S, two MACE minima connected by a verified saddle relaxed into one density-functional basin.
Reference relaxation likewise returned nine of ten selected model minima to their reference basins.
A directional test with two displaced reference calculations rejected 9 of 12 prospectively tested model instabilities.
Model agreement alone therefore cannot establish that predicted minima are distinct metastable states.

\end{abstract}

\section{Introduction}

The potential-energy surface (PES) of a material determines which structures it can adopt and how it transforms between them.
Universal machine-learned interatomic potentials (uMLIPs) now approximate density-functional PESs across much of the periodic table, making structure searches and transition calculations accessible at a fraction of the cost of reference electronic-structure methods\cite{batatia2025foundation,deng2023chgnet,chen2022m3gnet}.
These applications assume that the minima and saddles of a learned surface represent the states and transitions of the reference surface.
However, this assumption breaks down when distinct model minima relax into the same reference basin, a discrepancy we call \emph{local PES topology hallucination}.
Unlike errors in energies or equilibrium geometries, such hallucinations change which physical states are inferred.

Assessments of learned potentials have followed two complementary approaches.
Accuracy benchmarks have revealed a systematic softening of learned surfaces that limited additional reference data can substantially correct\cite{deng2025softening}, and phonon benchmarks show that accurate near-equilibrium energies and forces can coexist with an inaccurate harmonic response\cite{loew2025phonons}.
Landscape benchmarks instead map stationary points directly, revealing additional unphysical minima and incorrect transition-state connections in molecules\cite{carare2025landscape} and landscape discrepancies along symmetry-restricted coordinates in crystals\cite{parackal2026symmetry}.
The first approach probes the surface near known structures, whereas the second has been limited to molecules or predefined coordinates.

Neither approach tests whether curvature-sign errors shared across architectures create additional basins in periodic materials.
This gap is most consequential when agreement between models determines whether a reference calculation is performed.
Uncertainty estimates guide the selection of reference data in active learning\cite{peterson2017uncertainty,vandermause2020flare,zhang2020dpgen}, and heterogeneous ensembles can improve the ranking of prediction errors\cite{liu2025heterogeneous}.
Errors shared across architectures, however, are indistinguishable from correct predictions under any consensus criterion.
What is needed is a reference protocol that resolves predicted curvature signs without complete Hessians and determines whether each additional model basin corresponds to a distinct reference state.

Here we examine the released MACE, CHGNet and M3GNet potentials on periodic \(s\)- and \(p\)-block structures, using fixed-cell density-functional calculations with the Perdew--Burke--Ernzerhof (PBE) functional as a common reference.
Our analysis combines two complementary reference tests.
A directional reference test classifies a model-predicted instability as genuine or false from two displaced PBE calculations along the same direction at the same reference geometry, separating sign errors from shifts in model equilibrium positions.
Reference relaxation then starts from the model minimum reached along a predicted instability and tests whether a distinct basin survives on the PBE surface.
We find that model consensus is reliable for positive curvature but frequently wrong for negative curvature, and that the shared false instabilities are mode selective.
Most of the minima that the models reach along these instabilities relax back into existing basins of the PBE surface.
Together, these tests distinguish additional model minima from distinct metastable states supported by the reference calculation.

\section{Results}

\subsection{Agreement across architectures conceals false instabilities}

\begin{figure}[p]
\centering
\includegraphics[width=\textwidth]{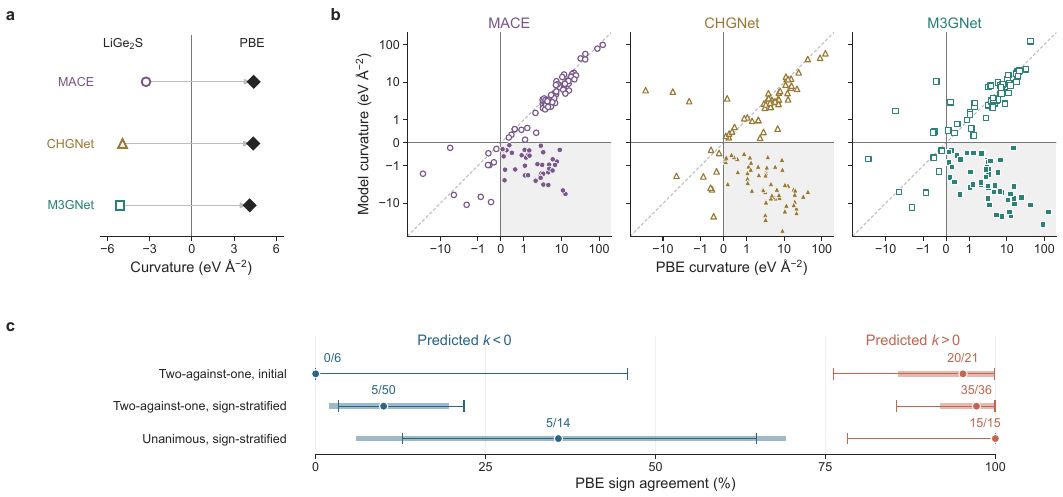}
\caption{\textbf{Model consensus fails for negative local curvature.} \textbf{a}, Curvatures \(k\) at the separately evaluated PBE-relaxed LiGe$_2$S geometry, each along that model's own selected direction. Open symbols denote model values and filled diamonds the PBE projections, with arrows connecting each pair. All three models predict negative curvature, whereas the PBE projections are positive. \textbf{b}, Model versus PBE directional curvatures for the same 115 valid directions in the sign-stratified sample. Shading marks negative model and positive PBE curvature, and symbols in this region are filled. PBE curvature is positive for 42 of 53, 57 of 65 and 53 of 60 model-negative directions for MACE, CHGNet and M3GNet, respectively. Dashed lines indicate equality. Symmetric-logarithmic axes are linear within \(\pm1\) eV~\AA{}\({}^{-2}\). \textbf{c}, PBE sign agreement for two-against-one predictions in the initial and sign-stratified samples, and unanimous predictions in the latter, grouped by predicted sign. Points give observed fractions and labels give correct/total directions. Thin capped and thick shaded intervals denote exact binomial and chemistry-family bootstrap 95\% intervals, respectively. Model colours and symbols are used throughout the paper, and blue and vermilion denote negative and positive curvature. The samples contain 42 and 115 valid directions.}
\label{fig:consensus}
\end{figure}

At the relaxed LiGe$_2$S reference geometry, each model predicted negative curvature along its own selected direction, whereas PBE curvature was positive along all three directions (Fig.~\ref{fig:consensus}a).
To test whether such disagreement is systematic, we compared signs along common candidate directions in two prospectively selected samples, using the same structure and direction for all three models and PBE.
PBE curvature was positive in 41 of the 42 directions of the initial sample, and an independent sign-stratified sample of 120 directions yielded 115 valid comparisons spanning positive and negative model agreement (Fig.~\ref{fig:consensus}b and Supplementary Note~2).

Agreement was far more reliable for positive than for negative predictions (Fig.~\ref{fig:consensus}c).
Positive two-against-one majorities matched PBE in 20 of 21 directions of the initial sample and 35 of 36 directions of the sign-stratified sample, giving 55 of 57 correct positive majorities against only 5 of 56 correct negative ones.
Even unanimous predictions in the sign-stratified sample showed this asymmetry, with 15 of 15 positive but only 5 of 14 negative predictions matching PBE, so agreement among all three architectures can endorse an instability that PBE contradicts.
The asymmetry persisted when whole chemistry families were resampled, with a 95\% bootstrap interval of 0.286--1.000 for the positive-minus-negative accuracy difference of these unanimous predictions.

\subsection{Curvature errors extend beyond uniform softening}

\begin{figure}[p]
\centering
\includegraphics[width=\textwidth]{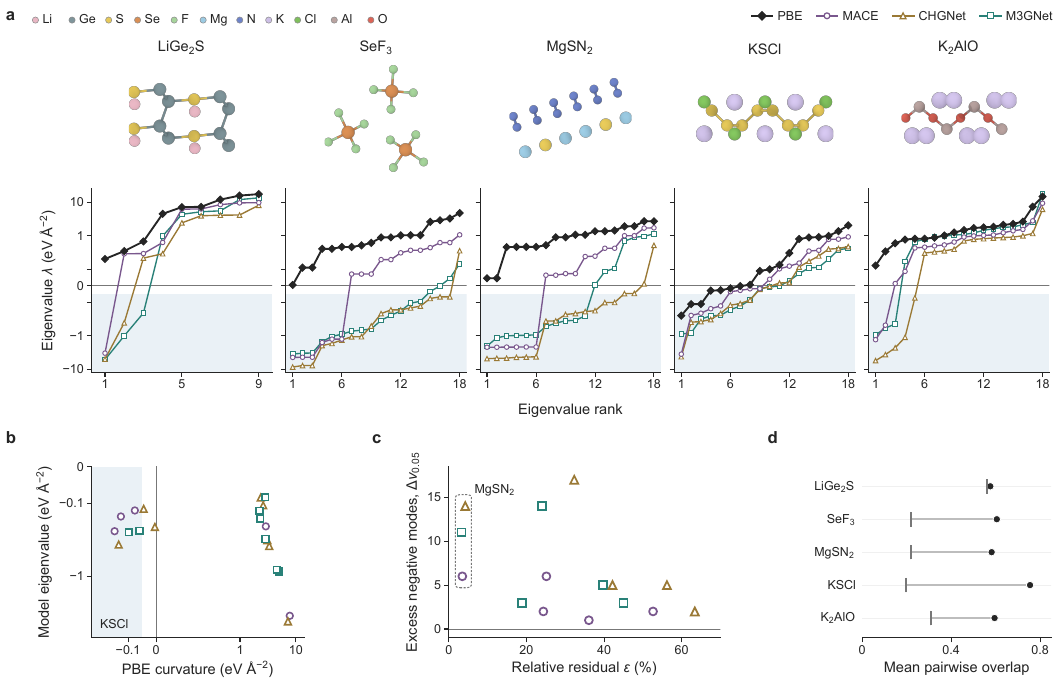}
\caption{\textbf{Curvature errors extend beyond uniform softening.} \textbf{a}, PBE reference structures (top), rendered with OVITO~\cite{stukowski2010ovito} at a common scale, and independently ordered, translation-reduced Hessian spectra at these geometries. SeF$_3$ is shown as its three SeF$_4$ units and the other structures as replicated cells. The spectra contain all nine eigenvalues for LiGe$_2$S and the lowest eighteen otherwise, a range that includes every eigenvalue below \(-0.05\) eV~\AA{}\({}^{-2}\) (blue shading). KSCl is the PBE-unstable reference. Lines join discrete eigenvalues ordered by rank, without eigenvector matching. \textbf{b}, Model eigenvalue against PBE curvature along the same direction for each KSCl model eigenvector with eigenvalue below \(-0.05\) eV~\AA{}\({}^{-2}\). Blue shading marks PBE curvature below \(-0.05\) eV~\AA{}\({}^{-2}\). \textbf{c}, Relative Frobenius residual \(\epsilon=\|\bar H_{\mathrm{model}}-\alpha\bar H_{\mathrm{PBE}}\|_F/\|\bar H_{\mathrm{model}}\|_F\) for translation-reduced Hessians \(\bar H\), after fitting a positive scalar \(\alpha\), against the number of model eigenvalues below \(-0.05\) eV~\AA{}\({}^{-2}\) in excess of PBE, \(\Delta\nu_{0.05}\), for all 15 case--model pairs. \textbf{d}, Observed mean pairwise overlap of negative residual subspaces (filled circles) and the 95th percentile of a selection-conditioned orientation distribution (vertical ticks, 10,000 draws). Random orientations preserve subspace ranks and the required projection onto the shared direction. Open symbols denote models and filled diamonds PBE. Symmetric-logarithmic axes in \textbf{a} and \textbf{b} are linear within \(\pm0.1\) eV~\AA{}\({}^{-2}\).}
\label{fig:mechanism}
\end{figure}

The shared false instabilities survived tightly converged PBE relaxation and displacement-step checks in five of seven candidate cases (Supplementary Notes~3 and~4).
Complete Hessians confirmed reference minima for LiGe$_2$S, SeF$_3$, MgSN$_2$ and K$_2$AlO, whereas model Hessians evaluated at the same PBE geometries contained negative modes (Fig.~\ref{fig:mechanism}a).
The unstable reference structure KSCl instead contained both false and genuine model-predicted instabilities (Fig.~\ref{fig:mechanism}b).
PBE curvature was positive along a direction that MACE and CHGNet predicted to be unstable, whereas a separate direction had negative PBE curvature in agreement with all three models (Supplementary Note~4).
Thirteen of the 21 model eigenvectors with eigenvalues below \(-0.05\) eV~\AA{}\({}^{-2}\) had positive PBE curvature, including the most negative eigenvector of each model, and the six with negative PBE curvature projected mainly onto the three negative PBE modes.
Each model reproduced the genuine instability of KSCl in at least one eigenvector yet placed its most negative curvature along a direction that is stable in PBE.

Positive scalar rescaling cannot correct such sign errors because it preserves the sign of every directional curvature.
For MgSN$_2$, relative Frobenius residuals after scalar fitting were only 3.3--4.2\%, yet all three models retained negative modes at the PBE minimum (Fig.~\ref{fig:mechanism}c).
The residuals also shared a directional component, with the subspaces of most negative residual curvature overlapping across architectures beyond the 95th percentile of a selection-conditioned orientation distribution in all five cases (Fig.~\ref{fig:mechanism}d).
Adding a fitted rank-five residual to the scalar approximation reduced the pooled root-mean-square error on directions excluded from the fit by only 16.5\% relative to scalar fitting, below the prespecified 50\% criterion (Supplementary Note~4).
The shared errors are mode selective, combining a common directional component across architectures with contributions that extend beyond five directions.

\subsection{Reference relaxation exposes additional model basins}

\begin{figure}[p]
\centering
\includegraphics[width=\textwidth]{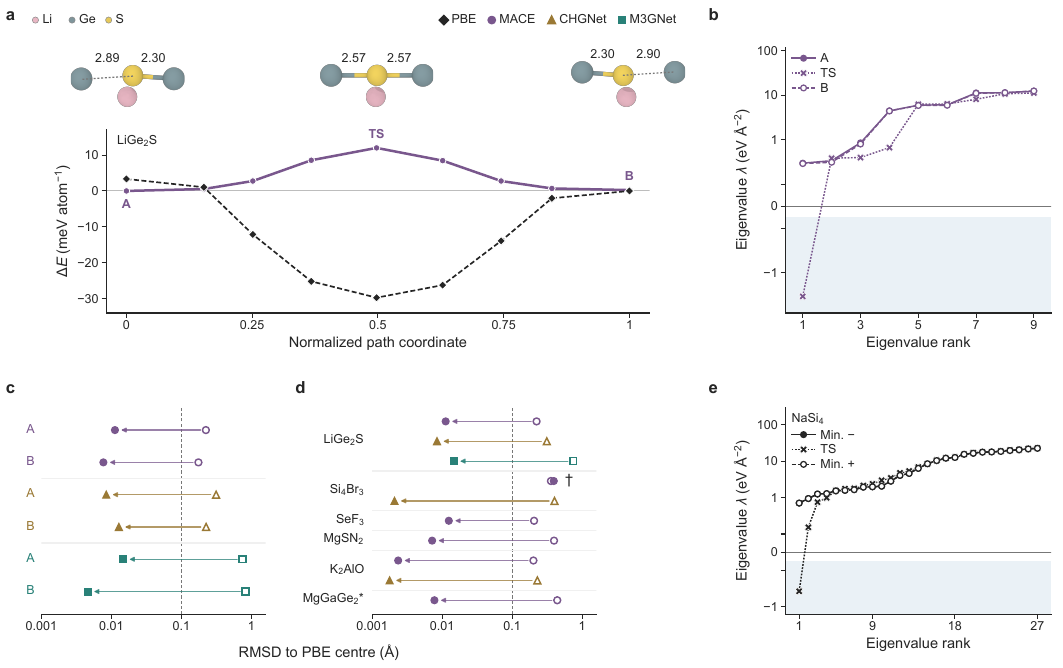}
\caption{\textbf{Reference relaxation removes distinct model-basin identities.} \textbf{a}, MACE and PBE energies at nine structures along the LiGe$_2$S MACE minimum-energy path, relative to each surface's lower path endpoint. Lines join calculated points, and the PBE single points evaluate the MACE path on the PBE surface. Above the corresponding path points, OVITO renderings of the S atom, its two Ge neighbours and Li at minima A and B and saddle TS share one projection and scale. Labels give three-dimensional minimum-image S--Ge distances, and dotted lines mark non-bonded contacts above 2.70~\AA{}. Periodic repeats are omitted. \textbf{b}, Complete translation-free MACE Hessian spectra of A, TS and B, with symbols encoding the stationary-point type. Maximum forces are 3.0, 2.7 and 4.5 meV~\AA{}\({}^{-1}\), below the 5 meV~\AA{}\({}^{-1}\) threshold. \textbf{c}, RMSD from the PBE centre for both basin representatives (A, B) of each architecture, at the model minimum (open) and after PBE relaxation (filled). \textbf{d}, The same comparison for ten selected model minima. Dashed lines in \textbf{c} and \textbf{d} mark the 0.1~\AA{} cutoff. The dagger marks the RMSD-distinct, energy-equivalent Si$_4$Br$_3$--MACE endpoint, and the asterisk the MgGaGe$_2$ sensitivity case. \textbf{e}, Complete PBE Hessian spectra for the NaSi$_4$ saddle and two distinct descent minima. Spectra in \textbf{b} and \textbf{e} are independently ordered without eigenvector matching. Their symmetric-logarithmic axes are linear within \(\pm0.1\) eV~\AA{}\({}^{-2}\), and blue shading marks eigenvalues below \(-0.05\) eV~\AA{}\({}^{-2}\). Each saddle has exactly one such eigenvalue.}
\label{fig:topology_complex}
\end{figure}

A false instability becomes consequential when the model surface develops a minimum along it.
For LiGe$_2$S, six signed displacements from the reference centre relaxed into two distinct basins on each model surface, and on the MACE surface a converged minimum-energy path (MEP) connected the two minima through a first-order saddle (Fig.~\ref{fig:topology_complex}a,b).
Hessian spectra established the minimum--saddle--minimum sequence, signed descents from the saddle recovered both endpoints, and the forward and reverse barriers were 11.99 and 11.75 meV atom\({}^{-1}\).
PBE relaxation nevertheless returned both model-basin representatives of each architecture to the same reference basin (Fig.~\ref{fig:topology_complex}c), and PBE energies sampled along the MACE path formed a single-well profile (Fig.~\ref{fig:topology_complex}a).
This many-to-one correspondence, together with the verified MACE saddle, distinguishes a topology hallucination from a shift of one equilibrium structure.

Reference relaxation returned nine of ten selected model minima to their respective reference basins (Fig.~\ref{fig:topology_complex}d).
These ten comprised one LiGe$_2$S minimum per model and seven minima from other case--model pairs (Methods), and we refer to the nine returning endpoints as \emph{phantom minima}.
The remaining MACE Si$_4$Br$_3$ endpoint lies beyond the root-mean-square displacement (RMSD) cutoff yet matches the centre crystallographically and in energy to within 1 meV atom\({}^{-1}\), so we classify it as RMSD-distinct and energy-equivalent (Methods).
The NaSi$_4$ control shows the opposite behaviour, retaining a genuine PBE saddle between two distinct PBE minima (Fig.~\ref{fig:topology_complex}e), so reference relaxation distinguishes phantom minima from genuine metastable states.

Both the false instabilities and the minima they produced varied among the tested structures and potentials.
Three larger-cell candidates retained false instabilities without developing additional stable model minima (Supplementary Note~5), and SevenNet-Omni predicted positive curvature for LiGe$_2$S, Si$_4$Br$_3$ and K$_2$AlO while retaining the genuine NaSi$_4$ instability\cite{park2024sevennet,kim2026sevennetomni}.
Each false instability marks a candidate for endpoint testing, and the endpoint search decides whether it creates an additional model basin.

\subsection{Directional reference tests distinguish false from genuine instabilities}

\begin{figure}[p]
\centering
\includegraphics[width=\textwidth]{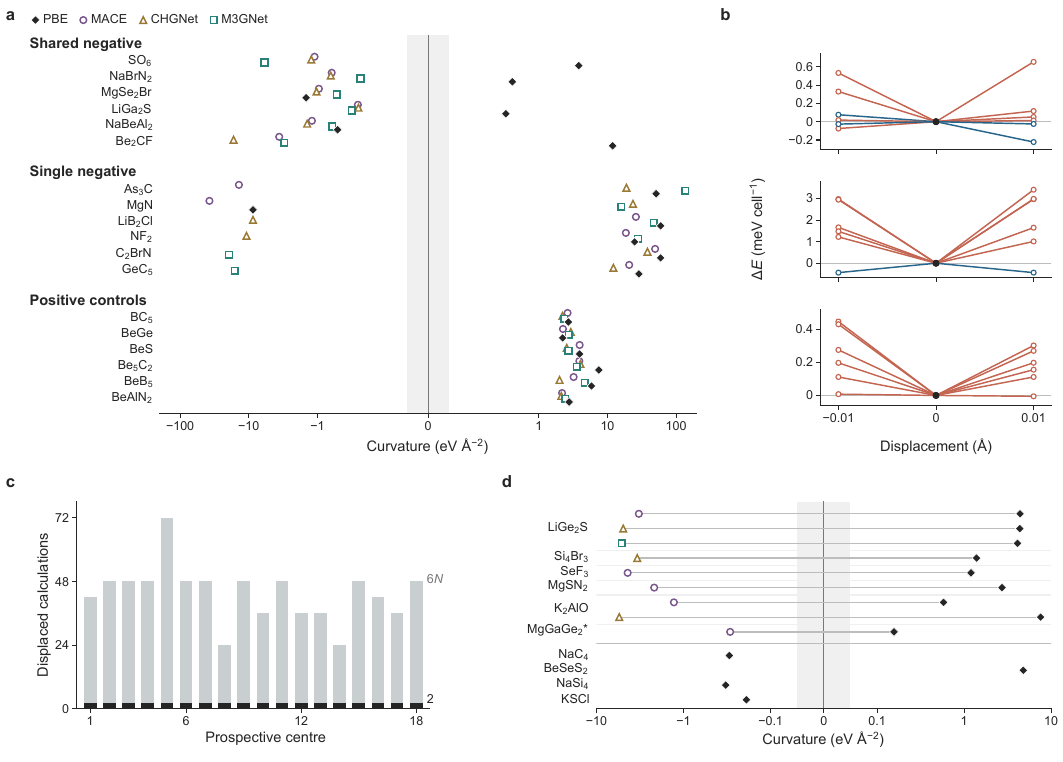}
\caption{\textbf{Directional reference tests distinguish false from genuine instabilities.} \textbf{a}, Model and force-derived PBE curvatures for 18 prospective directions, grouped by model predictions. Open symbols identify the three models and filled diamonds denote PBE. Small vertical offsets separate observations within each centre. \textbf{b}, PBE energies relative to each centre at displacements of \(-0.010\), 0 and \(+0.010\) \AA{}, grouped and vertically aligned as in \textbf{a}. Straight segments join calculated values, and centre points coincide. Blue and vermilion denote negative and positive PBE curvature, and energy scales differ between centres. \textbf{c}, Displaced-calculation counts for a complete Cartesian Hessian (\(6N\), for \(N\) atoms) and one directional test (2). Centres follow the top-to-bottom order in \textbf{a}. Totals are 780 and 36, excluding centre calculations and endpoint relaxations. \textbf{d}, Nine retrospective model/PBE pairs for originating directions associated with phantom minima, followed by four separate PBE controls. Values come from existing Hessians, and model symbols follow \textbf{a}. Lines connect paired values. The asterisk identifies the MgGaGe$_2$ sensitivity case. In \textbf{a} and \textbf{d}, symmetric-logarithmic axes are linear within \(\pm0.1\) eV~\AA{}\({}^{-2}\) and shading marks the \(\pm0.05\) eV~\AA{}\({}^{-2}\) ambiguity band.}
\label{fig:sparse_queries}
\end{figure}

Detecting a false instability requires only the reference curvature along the predicted direction, and a directional reference test obtains this curvature from PBE forces at two oppositely displaced configurations.
Curvature outside the ambiguity band classifies the direction as locally stable or unstable, and unresolved signs trigger additional displacements (Methods).
We evaluated the test prospectively on 18 previously unevaluated centres, one from each of 18 chemistry families.
Of twelve model-negative directions, nine were PBE positive and three were PBE negative, whereas all six model-positive controls remained positive (Fig.~\ref{fig:sparse_queries}a,b).
Every sign was resolved at the primary step, so the test used 36 displaced calculations where complete Cartesian Hessians would have needed 780, a reduction of 95.4\% overall and 95.8\% for the median structure (Fig.~\ref{fig:sparse_queries}c).
Two displaced PBE calculations per direction separated false from genuine instabilities in every prospective case.

Retrospectively, existing reference Hessians gave positive curvature along an originating model-negative direction for each of the nine phantom minima (Fig.~\ref{fig:sparse_queries}d), so a directional test would have rejected an originating instability of every phantom minimum.
A separate finite-displacement follow-up reached a distinct PBE basin 20.87 meV atom\({}^{-1}\) above its centre even though PBE had rejected the local instability along its originating direction.
Curvature describes the response to an infinitesimal displacement and endpoint relaxation the destination of a finite one, so the two tests together decide whether a displaced candidate leads to a distinct state.

\section{Discussion}

Our results show that a learned surface can match the reference Hessian closely and still contain minima and transitions that the reference surface lacks.
In LiGe$_2$S, MACE predicts two metastable states and a transition between them, whereas PBE relaxation returns both minima to one basin.
Force convergence and positive Hessian spectra certify such geometries as minima of the model surface, while their identity as distinct states belongs to the reference surface.
A crystal-structure search that accepts these minima would report more metastable structures than the reference surface supports.

Shared curvature-sign errors explain why model consensus fails as a safeguard.
Positive scalar rescaling preserves every curvature sign and cannot remove a false instability even when the overall Hessian fit is close, and the overlapping negative residual subspaces show that different architectures err along related directions.
Errors that all models share appear as confident predictions to any uncertainty estimate built on disagreement between models, and these directions are where reference evidence is needed most.
These findings complement studies of softening, vibrational response and landscape topology\cite{deng2025softening,loew2025phonons,carare2025landscape} by linking shared curvature-sign errors to the correspondence between model and reference basins in periodic crystals.

Directional reference tests make this evidence inexpensive to obtain.
Two displaced PBE calculations resolved the curvature sign of every prospective direction without a complete reference Hessian, and endpoint relaxation tested whether model minima remained distinct.
In one finite-displacement follow-up, a direction with positive PBE curvature still led to a distinct PBE basin, so screening workflows should pair the local test with endpoint relaxation before discarding a displaced candidate.

The present analysis covers selected local neighbourhoods of three released potentials against a fixed-cell PBE reference.
Sampling complete reference landscapes across chemical space would establish how often topology hallucinations arise, and extending the tests to cell strain, phonon dispersions, magnetic order and finite-temperature transition rates would carry them beyond fixed-cell local curvature.
The errors also vary between structures and models, and SevenNet-Omni avoided all three tested false instabilities.
The shared directional component identified here therefore offers a concrete target for training data and model design, and corrected potentials can be tested along these directions.
Structure-search and transition-path workflows built on universal potentials can apply both tests wherever a predicted instability or minimum enters a scientific conclusion, so that each reported metastable structure carries reference evidence of its identity.

\section*{Methods}

\subsection*{Models and reference calculations}

We used the released MACE-MP-0a medium, CHGNet MPtrj v0.3.0 and M3GNet-PES-MatPES-PBE-2025.2 potentials without retraining, evaluated with mace-torch 0.3.16, CHGNet 0.4.2 and MatGL 3.0.3, respectively.
MACE-MP-0a and CHGNet were trained on MPtrj data containing PBE and Hubbard-corrected labels, with additional magnetic-moment supervision in CHGNet, whereas the selected M3GNet was trained on MatPES-PBE\cite{batatia2025foundation,deng2023chgnet,kaplan2025matpes}.
The supplementary comparison used the MatPES-PBE checkpoint of SevenNet-Omni with SevenNet 0.13.0\cite{park2024sevennet,kim2026sevennetomni}, and all potentials were compared against a common PBE reference.

Reference calculations used VASP 6.6.0 with the projector augmented-wave (PAW) method and the PBE exchange-correlation functional\cite{kresse1996efficiency,kresse1999paw,perdew1996pbe}, a 520 eV plane-wave cutoff, Gaussian smearing of 0.05 eV and Gamma-centred reciprocal grids with spacing parameter 0.22 \AA{}\({}^{-1}\).
Cells were fixed throughout, and spin polarization, Hubbard corrections and symmetry operations were disabled in the primary calculations.
Electronic convergence thresholds were \(10^{-7}\) eV for the two sign-comparison samples and prospective directional tests and \(10^{-8}\) eV for reference relaxations and complete Hessians.
Reference relaxations were converged to a maximum atomic force of 0.005 eV~\AA{}\({}^{-1}\), and the analysis used only electronically converged calculations with finite energies and forces.
Supplementary Note~3 gives PAW choices and sensitivity tests, including spin-polarized calculations.

\subsection*{Direction selection and sign comparisons}

Periodic candidate structures were drawn from the ELEMENTA endpoint archive\cite{kairos2026elementa}, and binary and ternary \(s\)- and \(p\)-block compositions were selected using the structural, convergence and magnetic criteria in Supplementary Note~2.
Directions, model signs and selection thresholds were fixed before the target PBE evaluations.

For a unit Cartesian direction \(u\), the directional curvature is \(k(u)=u^{\mathrm T}Hu\), where \(H\) is the unweighted energy Hessian.
Model Hessians were calculated by central force differences at 0.010 \AA{} and symmetrized.
With \(N\) atoms and an orthonormal basis \(P\) perpendicular to the three rigid translations, we used the reduced matrix \(\bar H=P^{\mathrm T}HP\) of dimension \(3N-3\).
The two lowest modes of each model supplied candidate directions, and after removing near-duplicates we kept directions whose model signs persisted across 0.005, 0.010 and 0.020 \AA{} displacement steps.

At a centre configuration \(x\), reference curvature was estimated from opposite displacements of length \(h=0.010\) \AA{}:
\[
k_F(u)=-\frac{(F_+-F_-)\cdot u}{2h},
\qquad
k_E(u)=\frac{E_++E_--2E_0}{h^2}.
\]
Here, \(F_\pm\) and \(E_\pm\) are forces and total-cell free energies at \(x\pm hu\), and \(E_0\) is the free energy at \(x\).
The force-derived estimate was primary, and total-cell free energies provided a consistency check because finite-smearing forces differentiate the free energy.
We used an ambiguity threshold of 0.05 eV~\AA{}\({}^{-2}\) and the sample-specific consistency rules in Supplementary Note~3.

We analysed two-against-one and unanimous predictions separately, conditioning accuracy on the model majority sign, and calculated two-sided 95\% Clopper--Pearson intervals and bootstrap intervals from 10,000 resamples of whole chemistry families.
Supplementary Note~2 defines the samples, exclusions, estimands and statistical tests.

\subsection*{Hessian comparisons}

Selected discrepancies were retested after PBE relaxation to distinguish persistent curvature errors from differences in the reference geometry.
A discrepancy was counted as persistent when PBE curvature stayed positive at all three displacement steps, at least two models gave curvature at most \(-0.10\) eV~\AA{}\({}^{-2}\) and the tested direction had a squared projection of at least 0.50 onto each qualifying model's negative subspace.
Complete PBE Hessians used Cartesian central differences at 0.010 \AA{}, followed by symmetrization and translation removal.
Model Hessians were evaluated at the same PBE geometries so that model and reference curvatures refer to one configuration, although these geometries need not be stationary points of the model surfaces.

We counted eigenvalues below \(-0.05\) eV~\AA{}\({}^{-2}\) as thresholded negative inertia, \(\nu_{0.05}\), and reserved the Morse index for non-degenerate stationary points, where it counts negative eigenvalues after translations are removed\cite{milnor1963morse}.
Near-zero modes in SeF$_3$ and MgSN$_2$ were checked separately at two additional displacement steps.
For model \(M\), a non-negative scalar fit gave
\[
\alpha_M=\max\left\{0,\frac{\langle\bar H_{\mathrm{PBE}},\bar H_M\rangle_F}
{\|\bar H_{\mathrm{PBE}}\|_F^2}\right\},
\qquad
\Delta H_M=\bar H_M-\alpha_M\bar H_{\mathrm{PBE}}.
\]
The inner product \(\langle A,B\rangle_F=\sum_{ij}A_{ij}B_{ij}\) defines the Frobenius norm used in the relative residual \(\epsilon_M=\|\Delta H_M\|_F/\|\bar H_M\|_F\).
We compared the most negative residual subspaces using normalized pairwise projection overlap, with an orientation comparison that preserved their ranks and the directional persistence constraint.
A separate five-fold test fitted rank-five residual approximations to model curvatures at each centre, retaining the largest-magnitude eigencomponents of either sign, and each model was fitted separately with folds formed from directions within each centre.
Supplementary Note~4 gives both analyses, and Supplementary Note~1 gives the relevant local perturbation bounds.

\subsection*{Model minima and reference basin retention}

The endpoint search used four PBE minima from the persistent-Hessian comparison, Si$_4$Br$_3$ as a fifth primary case and MgGaGe$_2$ as a sensitivity case.
For every case--model pair whose lowest reduced Hessian eigenvalue was below \(-0.05\) eV~\AA{}\({}^{-2}\), relaxations started from the centre and from displacements of \(\pm0.025\), \(\pm0.050\) and \(\pm0.100\) \AA{} along that mode.
The 17 such pairs gave 119 starts, all relaxed at fixed cell using the fast inertial relaxation engine (FIRE)\cite{bitzek2006fire} with a 0.005 eV~\AA{}\({}^{-1}\) force threshold and at most 2,000 steps.
Endpoints from displaced starts became candidates when they lowered the model energy by more than 2 meV atom\({}^{-1}\) and lay more than 0.10 \AA{} RMSD from the reference centre, and the candidates were clustered separately for each case and model by species-constrained periodic RMSD.
The lowest-energy representative of each cluster was accepted as a model minimum when it was force-converged and either had a minimum reduced Hessian eigenvalue above +0.05 eV~\AA{}\({}^{-2}\) or qualified as a soft minimum.
A soft minimum had positive minimum reduced Hessian eigenvalues at all three displacement steps (0.005, 0.010 and 0.020 \AA{}), with force convergence at each evaluation.
Sixteen representatives were accepted, and the lowest-energy accepted minimum of each case--model pair gave ten endpoints for PBE relaxation.
The model-side selection rule was revised after initial Hessian inspection, before any endpoint PBE outcomes were known (Supplementary Note~5).

Each selected endpoint was relaxed at fixed cell with PBE under the reference conditions above and assigned to its reference basin when its RMSD from the centre was \(\leq0.10\) \AA{}, StructureMatcher\cite{ong2013pymatgen} matched the two structures and the endpoint-minus-centre energy difference satisfied \(\Delta E\geq-1\) meV atom\({}^{-1}\).
A returning model minimum was classified as a phantom minimum, and an endpoint above the RMSD cutoff with \(|\Delta E|<1\) meV atom\({}^{-1}\) was classified as RMSD-distinct and energy-equivalent.
The separate LiGe$_2$S comparison relaxed both model-cluster representatives per architecture, reusing three PBE relaxations and adding three (Supplementary Note~5).

\subsection*{Minimum-energy path calculations}

Model MEPs were calculated with climbing-image nudged elastic bands (CI-NEB) initialized by image-dependent pair-potential interpolation\cite{smidstrup2014idpp,henkelman2000cineb,larsen2017ase}.
Paths retained common endpoint atom ordering and wrapped periodic coordinate differences, and global translation was removed.
Primary paths used nine images including endpoints, with 15-image refinements where needed, and FIRE preconvergence and CI-NEB used force thresholds of 0.020 and 0.005 eV~\AA{}\({}^{-1}\), respectively.
The highest-energy interior image initialized dimer saddle refinement, and we accepted a refined saddle when its force was below 0.005 eV~\AA{}\({}^{-1}\) and its reduced Hessian had one eigenvalue below \(-0.05\) eV~\AA{}\({}^{-2}\) with all remaining eigenvalues above +0.05 eV~\AA{}\({}^{-2}\).
Its unstable mode also had to align with the path tangent, with an absolute inner product of at least 0.50 between the unit vectors, and signed displacements along the mode had to descend to the two distinct endpoints.
PBE single points at the model-path images trace a cut through the reference surface along the model path and compare the two surfaces on the same configurations.
Supplementary Note~6 gives refinement parameters, the NaSi$_4$ reference control and the descriptive polynomial fits.

\subsection*{Directional reference tests}

A model-negative direction was classified as false when its PBE curvature exceeded +0.05 eV~\AA{}\({}^{-2}\) and as genuine when it was below \(-0.05\) eV~\AA{}\({}^{-2}\).
The prospective test used one primary direction at each of 18 previously unevaluated centres from distinct chemistry families, comprising six shared-negative predictions, six directions negative in one model and six all-positive controls.
Each direction used the centre and two displacements at \(\pm0.010\) \AA{}, and ambiguous force curvatures or conflicting force and energy signs triggered additional evaluations at 0.005 and 0.020 \AA{}, with exact threshold ties left unclassified.
The retrospective comparison contracted existing reference Hessians along directions associated with the nine phantom minima, separately from the prospective test.

Without additional step-size checks, \(n_q\) queried directions require \(2n_q\) displaced calculations, compared with \(6N\) for a complete Cartesian Hessian of an \(N\)-atom cell.
The fractional reduction is \(1-n_q/(3N)\), excluding shared centre calculations and subsequent endpoint relaxations, and aggregate and median per-structure reductions were calculated separately.
Supplementary Note~7 gives the prospective outcomes, retrospective controls and independent endpoint follow-up.

\section*{Data availability}
Supplementary Information (\nolinkurl{Supplementary_Information.pdf}) is provided as an arXiv ancillary file alongside this preprint.
The structures, calculation settings, source data underlying Figs.~1--4 and Supplementary Fig.~S1, and analysis data generated in this study will be deposited in a public repository before formal publication and are available from the corresponding authors upon reasonable request.
The licensed PAW datasets cannot be redistributed, and the MACE-MP-0a, CHGNet, M3GNet and SevenNet-Omni potentials are available from their original releases\cite{batatia2025foundation,deng2023chgnet,kaplan2025matpes,kim2026sevennetomni}.

\section*{Code availability}
Custom code for the curvature comparisons, directional reference tests and endpoint analyses will be deposited in the same public repository before formal publication.

\bibliographystyle{pes-unsrtnat}
\bibliography{references}

\end{document}